\documentclass[]{elsart}

\usepackage{graphicx}

\newcommand{\vecb}[1]{{\vec{b}^{(#1)}}}
\newcommand{\vecr}[1]{{\vec{r}^{(#1)}}}
\newcommand{\vbe}[2]{{b^{(#1)}_{#2}}}

\begin{document}
\begin{frontmatter}
\title{Emergence of a complex and stable network in a model ecosystem
 with extinction and mutation}
\author[Tokita]{Kei Tokita\thanksref{KT}} and
\author[Yasutomi]{Ayumu Yasutomi\thanksref{AY}}
\address[Tokita]{Large-scale Computational Science Division, Cybermedia
 Center and\\
Interdisciplinary Computational Physics Group, Graduate School of
 Science, Osaka University, 1-32 Machikaneyama-cho, Toyonaka 560-0043, JAPAN}
\address[Yasutomi]{Department of Area Studies, Graduate School of Arts
 and Sciences, University of Tokyo, 3-8-1 Komaba, Meguro-ku 153-8902, Tokyo, Japan}
\thanks[KT]{URL : {\tt http://wwwcompy.phys.sci.osaka-u.ac.jp/\~{}tokita/}}
\thanks[AY]{E-mail address : {\tt yasutomi@ask.c.u-tokyo.ac.jp}}

%\vspace{2mm}
%{\small {\bf Running title}: Emergence of a complex ecosystem\hfill}

\begin{abstract}
We propose a minimal model of the dynamics of diversity --- replicator
equations with extinction, invasion and mutation.  We numerically
study the behavior of this simple model and show that it displays
completely different behavior from the conventional replicator
equation and the generalized Lotka-Volterra equation.  We reach
several significant conclusions as follows: (1) a complex ecosystem
can emerge when mutants with respect to species-specific interaction
are introduced; (2) such an ecosystem possesses strong resistance to
invasion; (3) a typical fixation process of mutants is realized
through the rapid growth of a group of mutualistic mutants with higher
fitness than majority species; (4) a hierarchical taxonomic structure
(like family-genus-species) emerges; and (5) the relative abundance of
species exhibits a typical pattern widely observed in nature. Several
implications of these results are discussed in connection with the
relationship of the present model to the generalized Lotka-Volterra
equation.
\end{abstract}
\end{frontmatter}

%\newpage

\section{Introduction}

The relationship between the complexity and stability of an ecosystem
has been one of the most fascinating topics in theoretical biology for
decades (Pimm 1991). In the 1950s and 1960s the proposition that highly
complex communities are more stable than simple ones was widely
supported (MacArther, 1955; Elton, 1958). However, this view was
challenged by theorists in the 1970s, who discussed the stability of a
community of species interacting randomly (Gardner \& Ashby, 1970; May,
1972, 1974; Roberts, 1974; Gilpin \& Case, 1976; Taylor, 1988a). The
most significant result was given by May (1972), who considered a
large-dimensional ecological equation with an $n$-dimensional random
interaction matrix $\{a_{ij} \}$ whose diagonal elements are $-1$ and
whose off-diagonal elements are assigned as Gaussian random numbers with
mean $0$ and variance $\alpha^2$ (with probability $C$) or $0$ (with
probability $1-C$). He found a kind of phase transition from stability
to instability in which the equilibrium solution corresponding to the
coexistence of all species becomes unstable if $\alpha>(nC)^{-1/2}$.  He
concluded that an ecological system cannot be stable if it is complex.
After May's work and the clear conclusions he reached, a number of
mathematical biologists attempted to explain the discrepancy between the
observed complexity of ecosystems in nature and the results of these
mathematical studies. Their works can be classified into two groups with
regarded to the methodology of assembling a complex and stable
ecosystem.

The first pioneering idea was formulated by Tregonning \& Roberts
(1979). They prepared a $50\times 50$ random interspecies interaction
matrix and checked the {\it feasibility} of the equilibrium solution of
a population equation. Here, an equilibrium solution is termed feasible
if all species have positive populations. In the case that the solution
under consideration was not feasible, they removed that species whose
equilibrium population was most negative.  Repeatedly applying this
procedure, they acquired a stable equilibrium of positive population
with considerably higher diversity than that predicted by May's theorem
[see also Roberts \& Tregonning (1980)]. Let us call their approach the
{\it eliminating approach}.  Although this approach can produce a
complex and stable ecosystem, the resulting diversity is in fact much
less than the initial diversity ($=50$).  Moreover, the biological
significance of their modeling was not very clear because
negative population is never achieved in nature and because the origin
of the random interactions was not discussed.  [See also Tokita \&
Yasutomi (1999) for a paleontological explanation of the formation of
such random interspecies interactions.]  Furthermore, according to the
analysis of {\it extinction dynamics} (Tokita \& Yasutomi, 1999), which
is a type of population dynamics similar to the elimination process, the
resulting diversity is independent of the initial diversity. This result
suggests that when the interactions are assigned randomly, a system with
arbitrarily large diversity cannot emerge through such an elimination
process. However it is still worthy of attention that
Tregonning \& Roberts (1979) reported that the most often-observed 
interspecies interaction of the survived subsystem was a prey-predator 
relationship, because it was recently demonstrated that coexistence of a
number of species can be achieved by an ecological model with
such prey-predator relationships even if the interactions are random
(Chawanya \& Tokita, 2002).

Other authors have used a second method, which is known as {\it
community assembly} (Pimm 1991). In this method, a model community is
assembled through a sequence of invasions of species. This type of
invasion is believed to reflect the natural tendency of species to
arrive sequentially rather than simultaneously. The method further can
be classified into two subgroups. The first trailblazing one is
represented by that of Post \& Pimm (1983). They prepare a pool of
species {\it ex ante\/} and then add them sequentially.  The properties
of the newly introduced species are then determined through a sampling
process that is completely independent of the properties of the species
existing in the community at the time of introduction.  A considerable
number of studies have followed this line (Robinson \& Valentine, 1979;
Yodzis, 1988; Taylor, 1988b; Drake, 1990; Case, 1990; Law \& Morton,
1996; Happel \& Stadler, 1998).  These models are realistic in the case
that we consider systems in which new species come from outside of the
existing community. We thus refer to this as the {\it invasion
approach}.

The second subgroup concerns a longer-term evolutionary process (Colwell
\& Winkler, 1984; Ginzburg {\it et al.}, 1988; Happel \& Stadler,
1998). In such models, a new species is introduced as a mutant of an
existing species. More precisely, parameters characterizing a new
species are generated by making appropriate changes to the parameters of
one of the existing species, rather than by sampling the parameter
values independently from a separate predetermined ensemble. We refer to
this method as the {\it mutation approach}.

Although these works of community assembly succeed in producing stable
communities whose numbers of species are larger than that predicted by
May's theorem, the coexistence of hundreds of species has not been
demonstrated. On the other hand, it is worthy of mention that in a
recent {\it mutation} approach (Drossel {\it et al.}, 2001),
considerably higher diversity has been realized than that by other {\it
mutation} approaches. They focused on influence of prey-predator
interactions to the evolution of food web structure. Consequently,
mutualisms and other interspecies relationships were not considered, and
``connectance'' (the percentage of non-zero values of interactions) was
low. In contrast, in the present study, we consider not only the
prey-predator type but also high connectance of interactions of mutualisms,
competitions and parasitisms.

%%
%% 2002/10/17
%% Referee #3 コメント2)に従い、「相互作用行列による種の定義には、
%% 生態学的な制限や構造が欠けている場合があること」に言及する文章
%% を追加
%%
In the present model, species are defined by interactions matrices,
following the preceding studies mentioned above. Such definition is,
however, still a controversial paradigm because interaction
coefficients are not real biological traits, like body size, but
rather summaries of these traits filtered through some interaction
rules. We neverthless adopt the interaction matrices because one of
the motivations here is to clarify the relationship between complexity
and stability of large ecosystems, i.e. {\it the paradox of ecology},
raised by May (1972). For that purpose, analyses of multidimensional
trait space would be essential, which is, in general, not a simple
task for the real biological traits. Here we only cite some approaches
on community assembly that does deal in real traits (Brown \& Vincent,
1992; Sasaki 1997; Geritz, {\it et al.}, 1998).

In this paper, we observe the behavior of a system based on the
replicator equation in which a species is eliminated when its
population becomes too small, and mutants or invaders are added
periodically. We consider the cases of three different rules governing
the generation of new species. We call these {\it invasive}, {\it
global} and {\it local}. The {\it invasive} rule can be compared
with the model of the {\it invasion approach}, and the {\it global}
rule with the {\it global mutation approach}.  The {\it local} rule is 
introduced here for the first time.

We will show that the {\it local} rule exclusively allows for the
assembly of a very large and complex ecosystem which contains hundreds
of species strongly interacting with each other. In addition, we observe
that the diversity and mutualism increase together. The
key point in our model is the introduction of a mutant species whose
relationships with the {\it dominant species} (with large populations)
are virtually the same as those of their parent species, but who can,
nevertheless, interact differently to {\it minor species} (with small
populations) as well as to those species which will come in the future.
The fitness of such a mutant is almost identical to that of its parent
when it is introduced, and hence it often behaves as a temporally 
neutral mutant. 

We will show that the minor mutants play a key role in
the emergence of a complex ecosystem.
% The process of speciation realized in this model presents a new view regarding {\it
% sympatric speciation}.  
This idea can be discussed in connection with {\it neutral molecular
evolution} (Kimura, 1983), which stresses the importance of the
accumulation of nondirectional mutations. Our theory is closely related
to the {\it ecological neutral evolution theory} for the abundance and
diversity of species in tropical rain forests and coral reefs proposed
by Hubbell (1979, 1997, 2001).  The size distribution of the relative
abundance of species obtained in our model is in good agreement with a
well-known distribution obtained from detailed observations of real
ecosystems.

We will also discuss the resistance to invasion of assembled ecosystems.
We find that an ecosystem assembled under the {\it local} rule has
perfect resistance to invasion if we keep it free from invaders which
are assembled according to the {\it invasive} rules for a sufficiently
long time.  Contrastingly, an ecosystem assembled under the {\it
global} rule cannot resist invaders.

This paper is organized as follows. In the next section we observe
the dynamics of the number of surviving species in our model. In
the third section, we discuss the resistance to invasion of the
assembled ecosystems.  The final section is devoted to discussion
of the relationship between the generalized
Lotka-Volterra equations and the replicator equations.

\section{The replicator equation with extinction and mutation}

\subsection{Model}

Here, we investigate a model based on the system of ordinary
differential equations 
\begin{equation}
\frac{{\rm d}x_i(t)}{{\rm d}t} = x_i(t)
      \left(\sum_{j=1}^{N_I} a_{ij}x_j(t)
     -
      \sum_{j,k=1}^{N_I} a_{jk}x_j(t)x_k(t)\right)\label{replicator_eq},
\end{equation}
called the {\it\, replicator equations
}(RE)(Hofbauer \& Sigmund, 1998), on a $N_I$-dimensional simplex:
\begin{equation}
\sum_{i=1}^{N_I} x_i(t) = 1\qquad\qquad (0\le x_i(t) \le 1).
                               \label{constraint}
\end{equation}
The variable $x_i$ denotes the population (or {\it relative abundance}) 
of the species $i$, and $N_I$ denotes the initial number of
species, that is, the initial value of the diversity.  The $(i,j)$-th
element of the matrix $A=(a_{ij})$ determines the effect of species
$j$ on the growth rate of species $i$.

It is known that the $N_I$-dimensional RE is topologically equivalent to
an $N_I-1$ dimensional generalized Lotka-Volterra equation (GLVE)
(Hofbauer \& Sigmund, 1998). We use the RE here, simply because they are
more tractable for a numerical simulation of dynamics of a large
ecosystem with high diversity and complex interactions. A comparison
between the GLVE and RE in the context of the dynamics of diversity is
given in the last section.

It should be noted that the RE may possess heteroclinic orbit even in
low dimensions ($N_I\ge 4$) (May \& Leonard 1975; Chawanya 1995 \&
1997).  When a heteroclinic orbit approaches a {\it saddle}, where some
species are `extinct', their population take extremely small values but
they never exactly reach zero, because the orbit is bound in the {\it
interior} of the simplex (\ref{constraint}). In the vicinity of such a
saddle, these population have such small values that they cause
underflow in naive numerical calculations.  However, if we continue a
calculation in such a region using the more sophisticated calculation
method of Chawanya (1995, 1997), we often observe some of these species
start to revive and cause the orbit to eventually leave for another
saddle, in particular for a system with high diversity and complex
interactions. This transition among saddles continues cyclically or
chaotically. The exponential approach of a population to zero and its
subsequent revival to order $O(1)$ plays an important role in
heteroclinic orbits. However, in the real world, such small population
cannot be realized.  
%In this sense, heteroclinic orbits have never been
%believed to be biologically significant. Such small population in the RE
%is also connected with the population explosion in the generalized
%Lotka-Volterra equation.

Considering this problem, we introduce a parameter $\delta\,(\ll 1)$,
the {\it extinction threshold}, into the dynamics described by
(\ref{replicator_eq}) and (\ref{constraint}) : At each discrete time
step, the population $x_k$ is set to zero if this quantity becomes less
than $\delta$. The population of the surviving species \{$x_i$\}$\,
(i\neq k)$ are then renormalized to satisfy $\sum_{i\neq k}x_i=1$. 
%%
%% Deleted by the suggestion of the referee #2 [2] on 2002/10/16
%%
%%This renormalization implies that the niche of an extinct species is
%%shared by the survivors. This assumption seems to be more realistic
%%than the diversity conservation presupposed by many models, in which
%%the niche of an extinct species is immediately taken over by an
%%absolutely new species.
%%
The diversity $N(t)$ decreases through the above described
process. We refer to this as the extinction dynamics (ED) model (Tokita
\& Yasutomi, 1999). With $\delta$, a stochastic effect on extinctions is
introduced into the fully deterministic RE, and hence, $\delta$ plays a
role analogous to that of random genetic drift. The introduction of
$\delta$ constitutes a finite size effect on the total population into
RE, because $\delta$ represents a minimum unit of reproduction for each
species, and its reciprocal $1/\delta$ represents to the permissible
population size.  That is, the
system explicitly possesses an energy constraint and is free from the
problem of population explosion often observed in the simulation of a
large-scale GLVE pointed out by Taylor (1988b). 
%% 
%% Deleted along the suggestion by the referee #2 [2] on 2002/10/16
%%
%% One of the noteworthy characteristics of the ED is this biomass
%% conservation introduced naturally and by definition.  Moreover, the
%% present model belongs to a class of the systems whose dimension or
%% number of degrees of freedom is a time-dependent variable (Tokita \&
%% Yasutomi, 1999). These points have not been considered in studies of
%% the GLVE.

In contrast to the typical ED simulation with a large $N_I$, in the
present simulation, we start from only one species (i.e. $N_I=1$),
following Ginzburg {\it et~al.}(1988). 
The initial intraspecies
interactions is set to $-v\, (<0)$, which is one of the parameters in the
present model. A simulation proceeds by first computing the ED for a
fixed number ($T$) of time steps. This is followed by the introduction of
a new species. This procedure is then repeated many times.
We set $T=900$ in our simulations.
Hereafter, we refer to $T$ time steps as {\it one period\/}, and we use
$\tau$ to count the number of periods.  At the end of each period, we
select one parent species, $j$, among the existing species [$1 \leq j
\leq N(\tau)$] with a sampling probability proportional to the population
$x_j$. Then we produce a new species $k\, [=N(\tau)+1)]$ from
the parent species $j$, copying $\{a_{ji}\}$ and $\{a_{ij}\}$ to
$\{a_{ki}\}$ and $\{a_{ik}\}$ for all $i$, with some variations
applied in the manner described below. We then set the new species'
population as $x_k=\zeta$
($\zeta > \delta$) and normalize all $x_i$ in order to maintain the condition $\sum x_i=1$.
Thus completing the introduction of the new species, we then return to
the ED, and the entire process is repeated. 

\begin{figure}[t]
\begin{center} 
\includegraphics[angle=-90,width=0.55\textwidth]{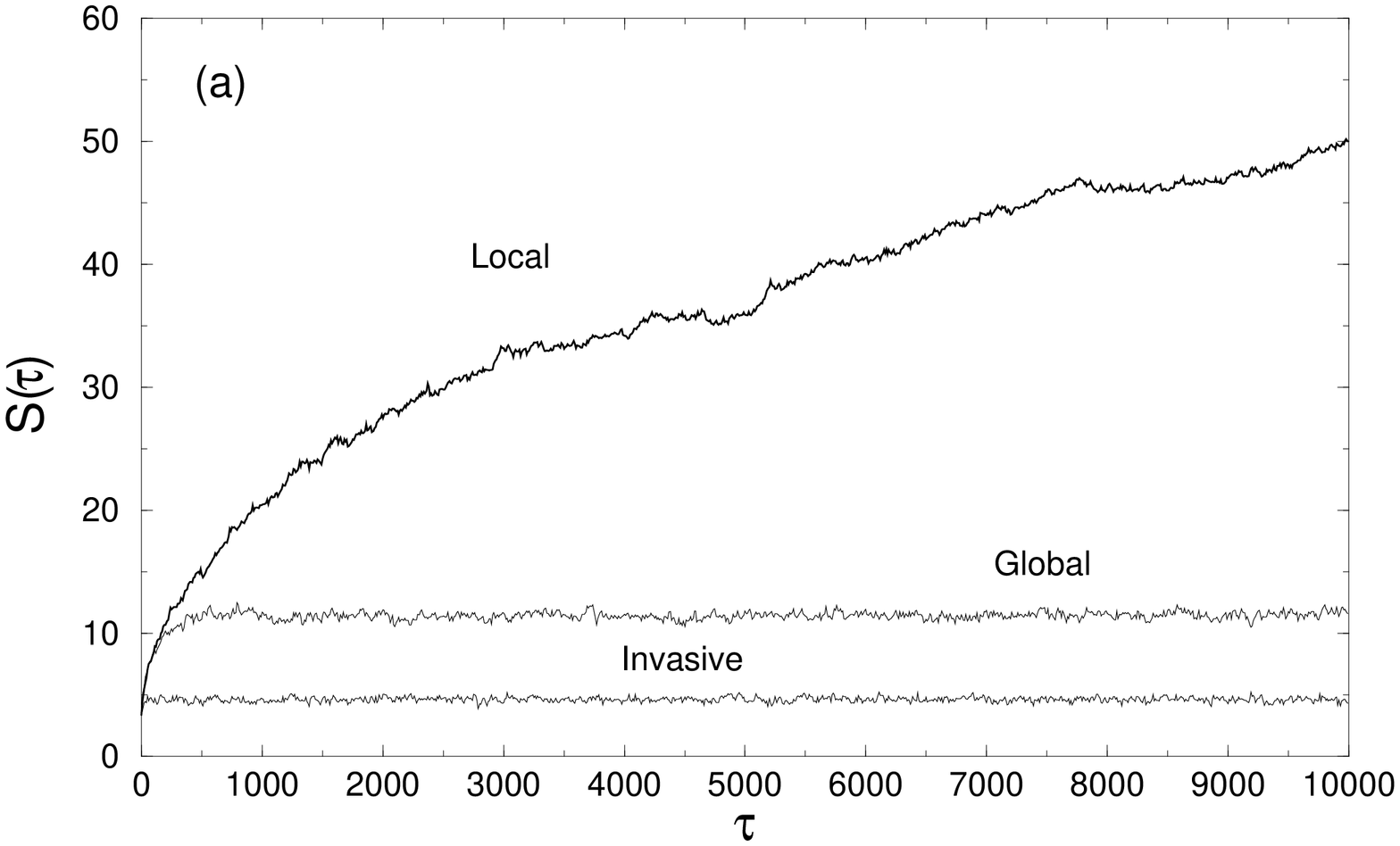}
\includegraphics[angle=-90,width=0.43\textwidth]{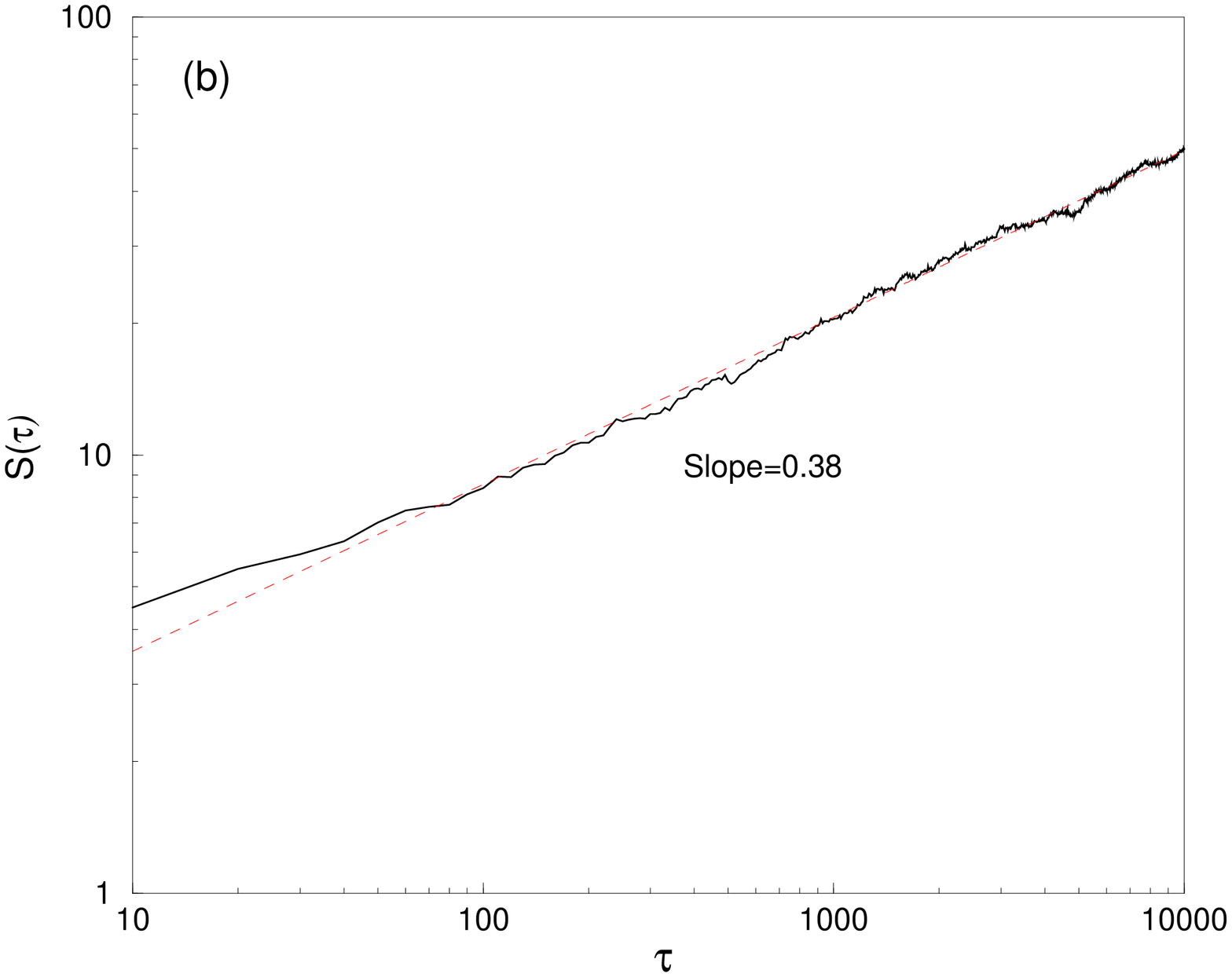}
\caption{\label{diversity} Only the {\it local} mutation
	   rule allows the number of {\it observable} species
	   to continue to increase. (a) The horizontal
	   axis $\tau$ represents the time measured with respect to the
	   interval between `mutations'. The vertical axis
	   represents the number of {\it observable} species $\bar{S}(\tau)$.
	   (b) The log-log plot of $\bar{S}(\tau)$ under the {\it
	   local} rule from (a) and the function with which it is fitted,
	   $\tau^{0.38}$ (dashed curve).
	   The parameters used here are $\delta=\exp(-11)$, $\zeta=\exp(-10)$,
	   $\xi=\exp(-9)$, $v=1.0$ and $T=900$. }
\vspace{5mm}
\end{center}
\end{figure}

As we discussed above, we consider three different rules for the
variations of the interactions $\{a_{ki}\}$ and $\{a_{ik}\}$ for the new
species: {\it invasive}, {\it global} and {\it local\/}. In the case of
the {\it invasive\/} rule, we add a new species whose interaction
coefficients are assigned randomly and are independent of all existing
species (including its parent). In the case of the {\it global\/} rule,
we produce a mutant whose interaction coefficients are all only slightly
different from those of its parent species. In the case of the {\it
local\/} rule, we introduce a mutant that is identical to its parent
species, apart from its interaction with one particular surviving
species. While the {\it invasive\/} (Robinson \& Valentine, 1979;
Yodzis, 1988; Happel \& Stadler, 1998) and {\it global\/} (Ginzburg {\it
et al.\/}, 1988; Happel \& Stadler, 1998; Drossel {\it et al.}, 2001)
rules have been employed in previous studies, the {\it local\/} rule is
examined for the first time in the present study. In terms of
theoretical population genetics, the {\it invasive} rule corresponds to
the ``house-of-cards'' approximation (Turelli, 1983).  The details of
these rules are as follows.
\begin{itemize}
\item {\it Invasive\/}: We assign the values of $\{a_{ki}\}$ and
      $\{a_{ik}\}$ for all $i\,(\neq k)$ as Gaussian random numbers 
      (mean 0 and variance 1). This implies that the mutant species is
      independent of its parent species.

\item {\it global\/}: Quantities $\{\upsilon_{ki}\}$ and
      $\{\upsilon_{ik}\}$ representing Gaussian random noise
 with mean $0$ and
variance $1$ are independently added to $\{a_{ki}\}$ and
$\{a_{ik}\}$ for all $i$ according to
\begin{equation}
 \left( 
  \begin{array}{c}
   a_{ki} \Leftarrow \gamma a_{ki} + (1-\gamma) \upsilon_{ki}\\
   a_{ik} \Leftarrow \gamma a_{ik} + (1-\gamma) \upsilon_{ik}
  \end{array}
\right.
\end{equation}
for all $i$, where $\gamma$ ($0<\gamma<1$) denotes the strength of the
noise. In our simulations, this value was set to $0.9$.

\item {\it local\/}: We select a species $l$ whose population
is non-zero and change only the relationship between the mutant
species and species $l\,(\neq k)$. Thus, the effects of a mutation are
concentrated on only two elements, $a_{kl}$ and
$a_{lk}$. We assigned the values of these two elements using Gaussian random
numbers with mean $0$ and variance $1$.
\end{itemize}
In all cases, the intraspecies interaction $a_{ii}$ takes the value $-v$
for each species $i$.  Finally, we introduce a threshold $\xi$ ($\xi >
\zeta > \delta$), and we consider a species to be {\it unobservable} if
its population $x_i$ is smaller than $\xi$. Since $\xi$ is
larger than $\zeta$, a mutant species is not {\it observable} until its
population increases and becomes greater than $\xi$. The
function $S(\tau)$
denotes the number of the {\it observable} species existing at the end of the
$\tau$th period. We regard $S(\tau)$ as the diversity in this model.  It
should be noted that, in general, there appear much more {\it
unobservable} species than {\it observable} species although the
population of the latter is much larger than that of the former.

\subsection{Results}

\begin{figure}[t]
\begin{center}
\includegraphics[angle=-90,width=0.55\textwidth]{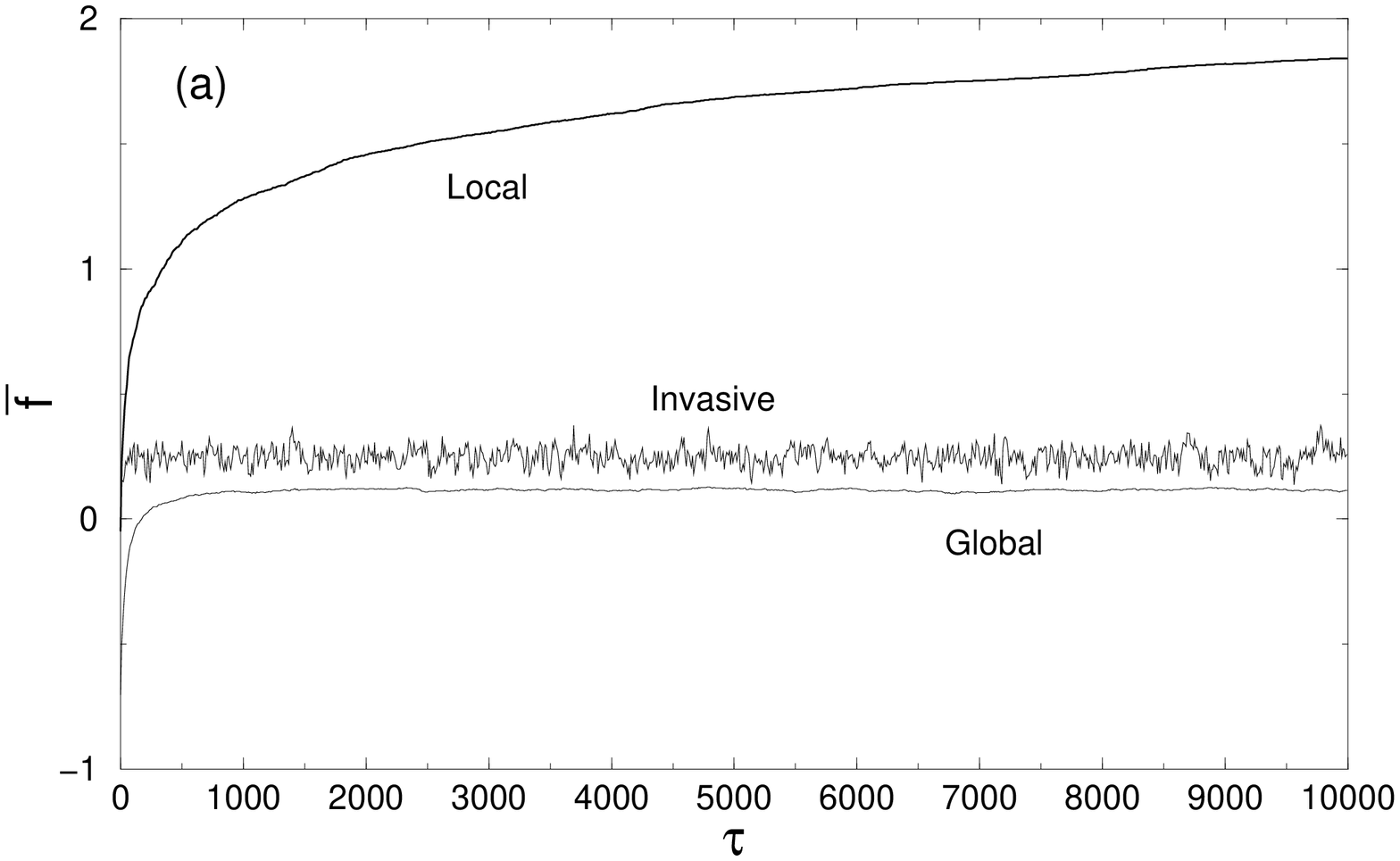}
\includegraphics[angle=-90,width=0.43\textwidth]{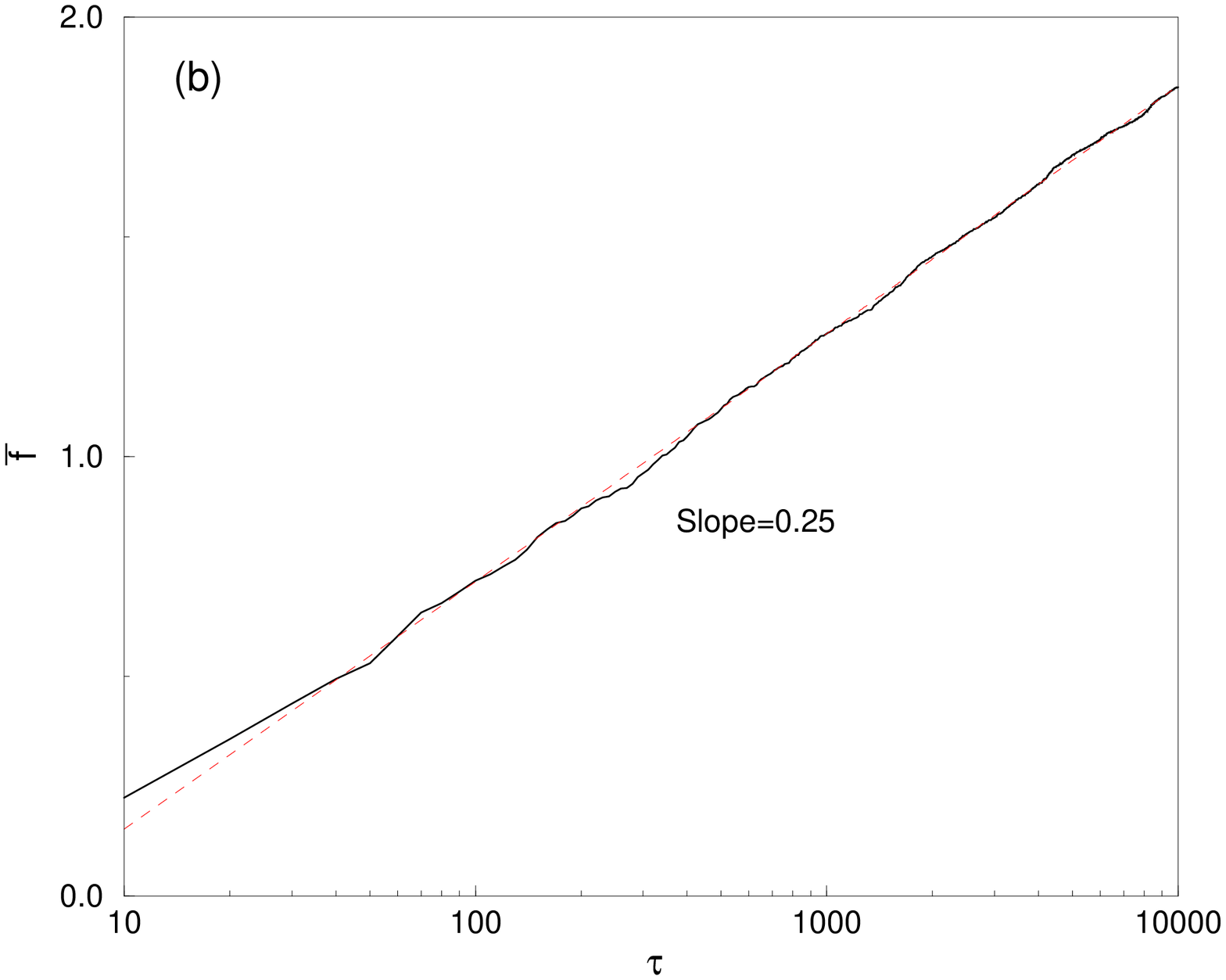}
\caption{\label{AveSf} (a)The average fitness,
	   $\bar{f}(t)\equiv\sum_{j,k=1}^{N(t)}a_{jk}x_j(t)x_k(t)$.
	   All curves were obtained from an average over $50$ samples of
	   independent simulations with different series of random numbers.  (b) The log-normal
	   plot of $\bar{f}$ under the {\it local} rule from
	   (a) and the function with which it is fitted,
	   $\log(\tau^{0.25})$ (dashed line).}
\end{center}
\end{figure}

Figure \ref{diversity} (a) displays the development of $\bar{S}(\tau)$
for each of the three rules, where $\bar{S}(\tau)$ is defined as
$S(\tau)$ averaged over independent simulations with different series of
random numbers.  We set the initial conditions as $x_1=1$ [that is,
$N(0)=S(0)=1$] and $a_{11}=-v$. This figure shows that diversity grows
most under {\it local} rule.
The diversity $\bar{S}(\tau)$ under the {\it local} rule
increases to approximately $50$ at $\tau = 10,000$, and as can be
inferred from this plot, it continues to increase beyond this value as
more mutants are added.  We averaged the diversity over $50$ simulations
and found that this averaged value obeys
$\bar{S}(\tau)\sim\tau^{0.38}$ [Figure \ref{diversity}(b)], which should be
compared with the square root law $S(\tau)\sim\sqrt{\tau}$ in the
coinfection model (May \& Nowak, 1995).

Figure \ref{AveSf} (a) displays the development of the average
fitness under the three rules. As in the case of the diversity, it is
seen here that only the {\it
local\/} rule allows for the average fitness to continue to increase.
Since the average fitness $\bar{f}=\sum_{j,k=1}^{N(\tau)}
a_{jk}x_j(\tau)x_k(\tau)$ is a kind of average of the interaction
coefficients, we can regard this quantity as an index of mutualism. The
results here therefore indicate that the diversity and the level of the
mutualism increase together. It is interesting that the average fitness does
not display power-law behavior, but rather increases 
much more slowly: $\bar{f}\sim c\log(\tau)\, (c\sim
0.25)$ (Figure \ref{AveSf}(b)). This is analogous to the  slow increase
of the diversity observed in a GLVE model of
super-infection in a host-parasite system (May \& Nowak, 1994).

\begin{figure}[t]
 \begin{center} 
\includegraphics[angle=-90,width=0.48\textwidth]{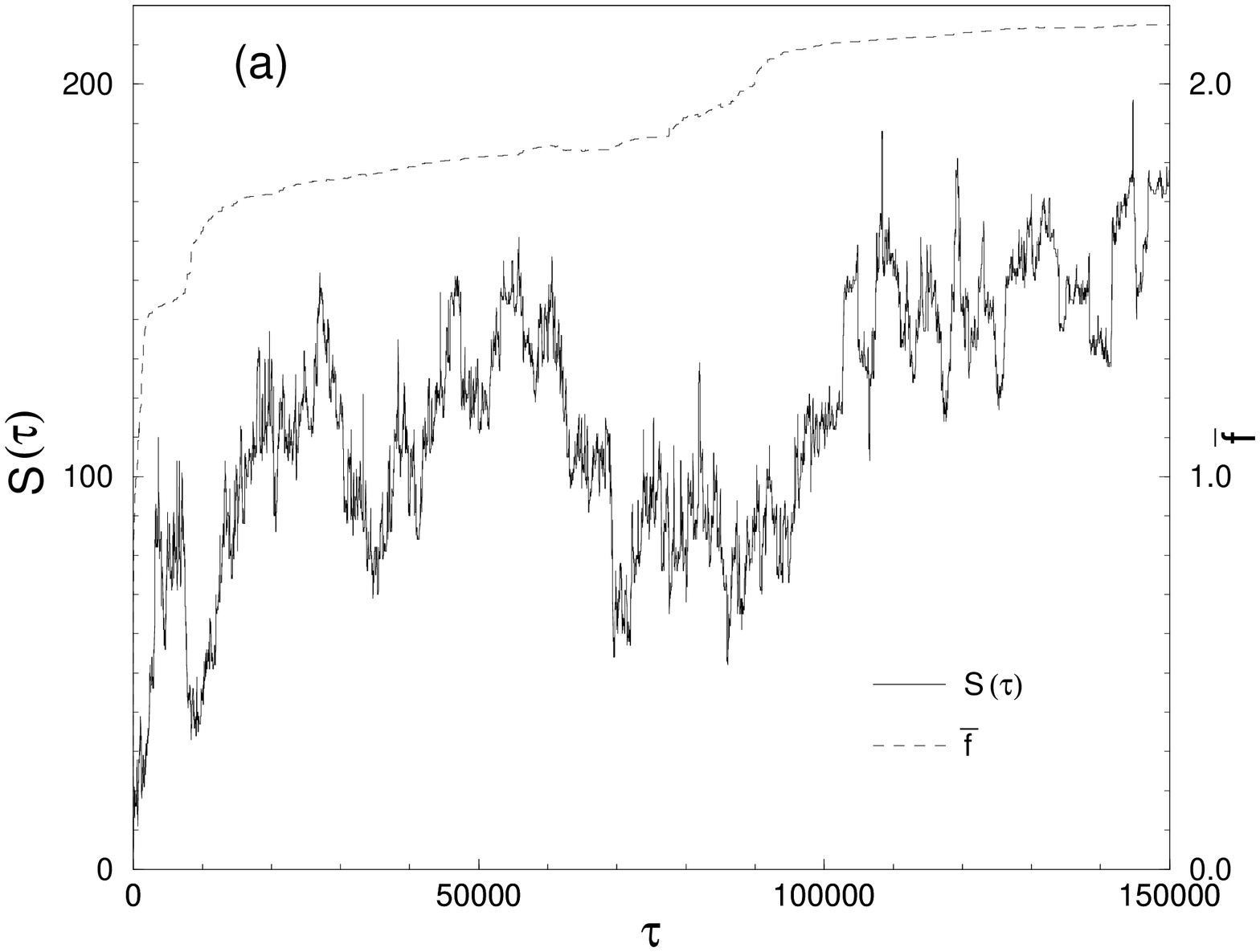}
\includegraphics[angle=-90,width=0.48\textwidth]{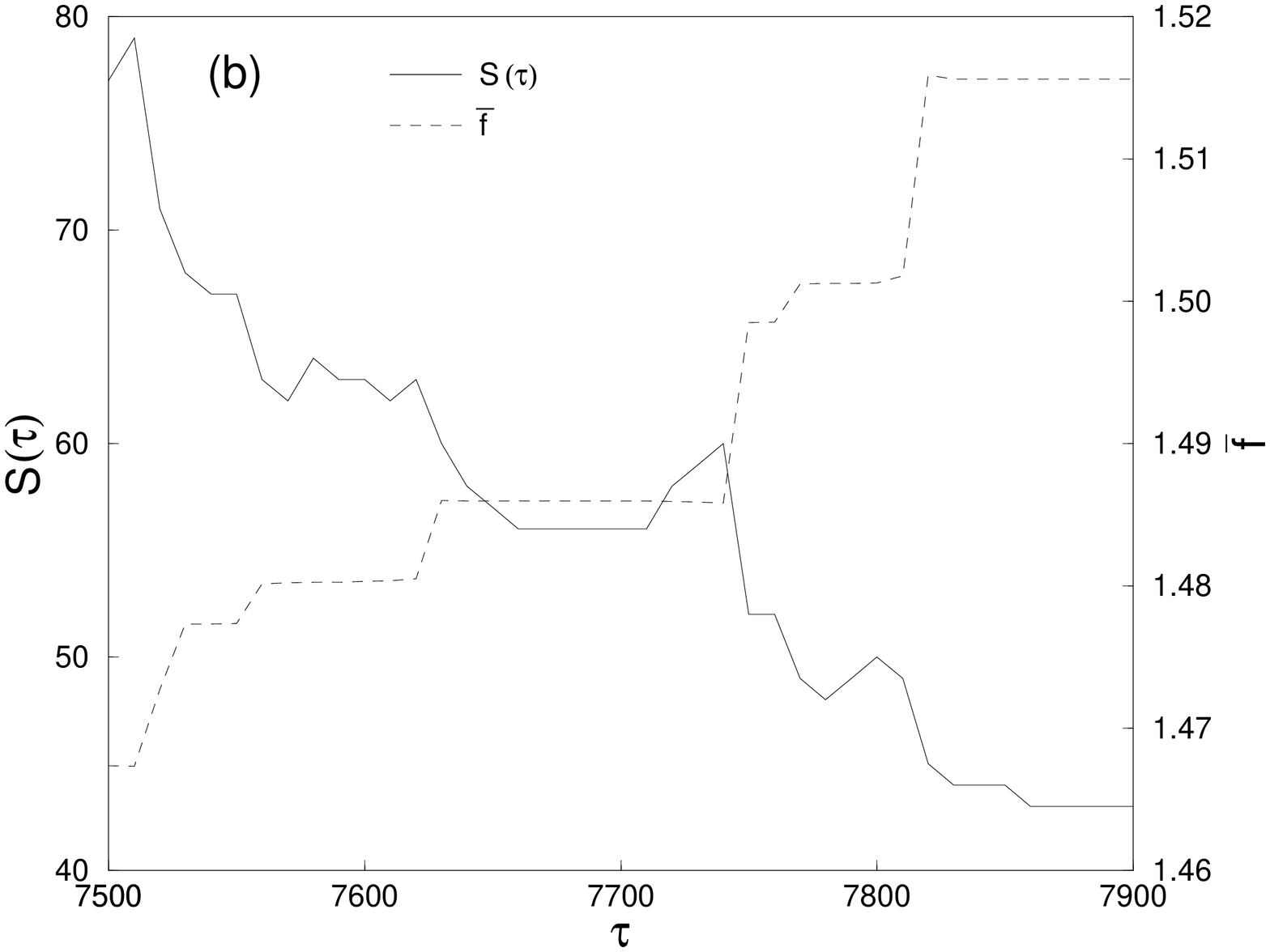}
\caption{\label{Sftau} (a) Development of the diversity $S(\tau)$ and
	   the average fitness $\bar{f}$. This figure shows the result for
	   a single sample under the {\it local} rule. 
       (b) An enlarged view of (a) for $\tau=7500 - 7900$. Here
       $\delta=\exp(-11)$, $\zeta=\exp(-10)$, $\xi=\exp(-9)$, $v=1.0$ and $T=900$.
}
\vspace{5mm}
 \end{center}
\end{figure}

Figure \ref{Sftau}(a) plots the development of $S(\tau)$, where the
largest number of species observed for this case is $196$ at $\tau =
144,700$.  In the plots of Figures \ref{Sftau} (a) and (b),
only one sample for the development of each $S(\tau)$ and $\bar{f}$ is
used. By contrast, the plots in Figure \ref{diversity} represent
averages over $50$ samples. From Figure \ref{Sftau}(a), we wee that the
diversity does not increase monotonically, and repeated avalanches of
extinction followed by rapid recoveries are observed. Figure
\ref{Sftau}(b) gives an enlarged view of the time evolution plotted in
Figure \ref{Sftau}(a) from $\tau=7,500$ to $7,900$ in order to clarify
the relationship between $S(\tau)$ and $\bar{f}$. Here we observe a
clear negative correlation; that is, the average fitness drastically
increases during an avalanche of extinction. This behavior is typically
observed in extinction dynamics (Tokita \& Yasutomi, 1999).  However,
the average fitness does not increase during the time that the diversity
increases. This implies that extinction plays a more important role than
speciation in the increase of the average fitness.  This is because
large extinction events swiftly exterminate less mutualistic species and
produce a more mutualistic network of species. The refined network then
prepares for a rapid recovery of the diversity in the aftermath of the
extinction.

In the case that $S(\tau)$ is defined with just one sample, its
fluctuation suggests simultaneous occurence of extinction and speciation
of multiple species. On the other hand, $\bar{f}(\tau)$ develops
relatively smoothly, because the definition of $\bar{f}$ is a kind of
average of the interactions $\{a_{ij}\}$ over $\{x_i\}$, and $x_i$ at
the time that the $i$th species appears or extincts is very small
($\delta$ for extinction and $\zeta$ for speciation).

\begin{figure}[t]
\begin{minipage}{0.52\textwidth}
  \includegraphics[angle=-90,width=\textwidth]{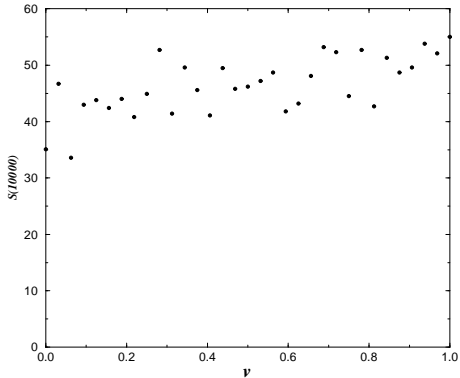}
\end{minipage}
\hspace{0.02\textwidth}
\begin{minipage}{0.46\textwidth}
  \caption{The {\it observable} diversity $\bar{S}(\tau)$ at
		   $\tau=10,000$ displayed as a function of $v$ under the
		   {\it local} rule.
		   $v$ represents the value of diagonal elements of the interaction
	   matrix (the intraspecies interaction), while the off-diagonal
		   interspecies interactions are assigned as Gaussian random
	   numbers with mean $0$ and variance $1$. Here
	   $\delta=\exp(-11)$, $\zeta=\exp(-10)$, $\xi=\exp(-9)$
	   and $T =100$. Each dot corresponds to an average over
	   10 samples of independent simulations.}
  \label{var} 
\end{minipage}
\vspace{5mm}
\end{figure}

Figure \ref{var} plots the dependence of $\bar{S}$ on the ratio of the
value of the diagonal elements, $v$, and the variance of the Gaussian
random numbers ($=1$) that are used to assign the interaction
coefficients of the mutant in the case of the {\it local\/} rule. We
see that $\bar{S}$ depends
only weakly on $v$. We should note that May (1972) showed that the
internal equilibrium point becomes unstable when this value becomes
less than $1$ if the system is assembled randomly at one time, rather
than through the periodic introduction of mutants, as done in our study.
The strength of the interactions
between species affects the stability of the ecosystem if it is
assembled at one time, but it does not do so when
the ecosystem gradually develops for a long time on the evolutionary
time scale.

\begin{figure}[t]
\begin{minipage}{0.48\textwidth}
  \includegraphics[angle=-90,width=\textwidth]{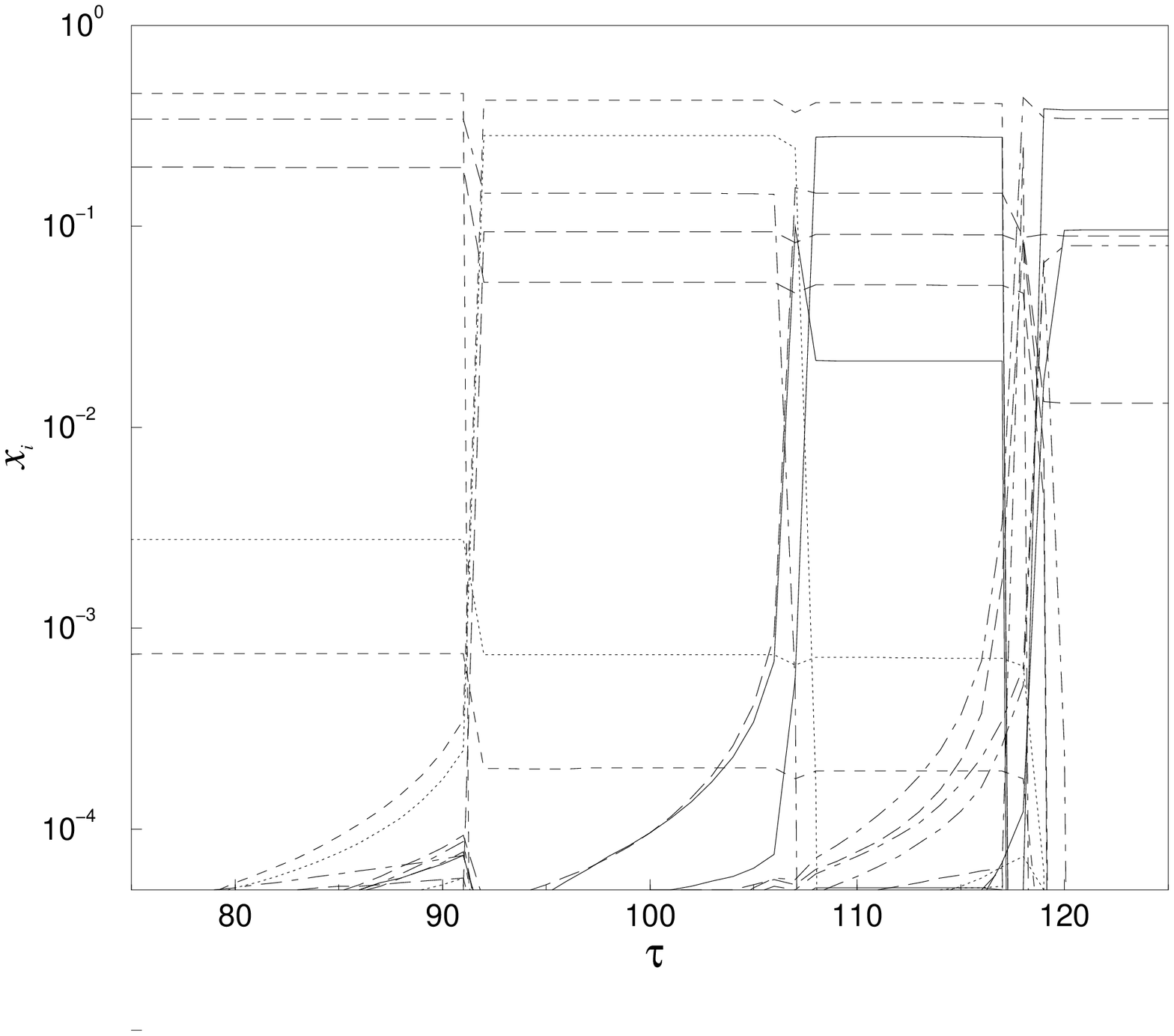}
  \caption{A few mutant species grow more
	   rapidly than exponentially, and they enter into the
	   {\it observable} ecosystem at virtually the same time. Here
	   $\delta=\exp(-11)$, $\zeta=\exp(-10)$, $\xi=\exp(-9)$, $v=1.0$ and
		   $T =900$.}
  \label{Local_Keiro} 
  \end{minipage}
  \hspace{0.04\textwidth}
  \begin{minipage}{0.48\textwidth}
  \includegraphics[angle=-90,width=\textwidth]{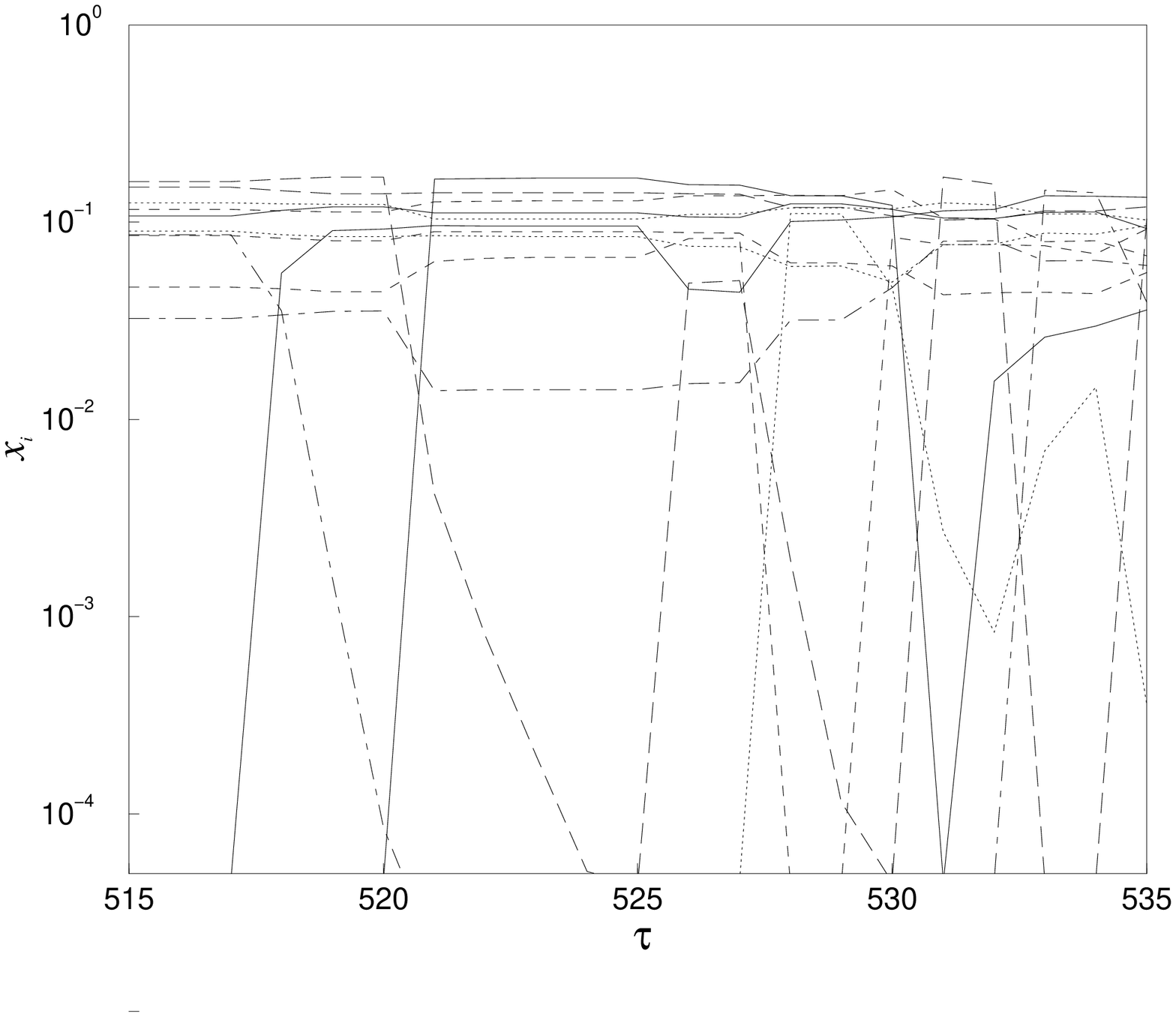}
  \caption{Mutant species produced
	   through the {\it mutational} rule grow less rapidly than
	   exponentially. After entering the ecosystem, they usually come
		   to replace their parent species. Here
	  $\delta=\exp(-11)$, $\zeta=\exp(-10)$, $\xi=\exp(-9)$,
	   $v=1.0$ and $T =900$.}
  \label{Global_Keiro} 
 \end{minipage}

 \vspace{5mm}
\end{figure}

Why does the diversity $S(\tau)$ increase so drastically only in
the case of the {\it local\/} rule?  To answer this question,
let us consider Figure \ref{Local_Keiro}, which gives an
example of our numerical integration. This figure shows that a few
species enter the existing ecosystem more rapidly than
exponentially and virtually simultaneously.  This kind of rapid
population growth results from the fact that the relationship
between these mutant species is more mutualistic than those between
the dominant species. When the relationship between mutants $p$
and $q$ is more mutualistic than those between the dominant species,
and this relationship therefore makes these mutants ``fitter''
than the average species, their population increase. The
growth of the population of species $p$ thus increases the
fitness of species $q$, thereby enhancing its growth rate, and
hence increasing the fitness of $p$. These {\it unobservable} species
constitute a new mutualistic {\it unobservable} subsystem and eventually merge
into the {\it observable} existing ecosystem. The implication of this
behavior is that the formation of a mutualistic network is more
effective as a invasion strategy than competition or even
exploitation under the pressure of natural selection, because a
mutualistic invasion has the abovementioned positive feedback effect,
while the exploitation involves a negative feedback (as the more one
species preys on another, the less abundant this prey becomes).

In order to form such a subsystem, mutants must wait for the arrival
of partners. In order to avoid extinction while waiting, such a
mutant must have virtually the same fitness as its parent
species. More precisely, they must have the same interaction coefficients
with respect to the dominant species as their parents. If this is not
the case, they will independently invade the existing ecosystem,
replacing their parent species, or they will simply be repulsed. Of
course, the {\it invasive\/} and {\it global\/} rules rarely produce
such ``sleeping'' mutants, because in general in these cases all interactions of a given mutant
will be different from those of their parents. Figure
\ref{Global_Keiro} gives an example of our numerical
integration for the case of the {\it global\/} rule. Here a new species
invades independently, replacing an existing species, and we can observe
no faster-than-exponential population growth. Only the {\it
local\/} rule can produce such mutants, and this is the reason why
only it can yield a remarkable increase in the diversity.

\begin{figure}[t]
\begin{minipage}{0.48\textwidth}
  \includegraphics[angle=-90,width=\textwidth]{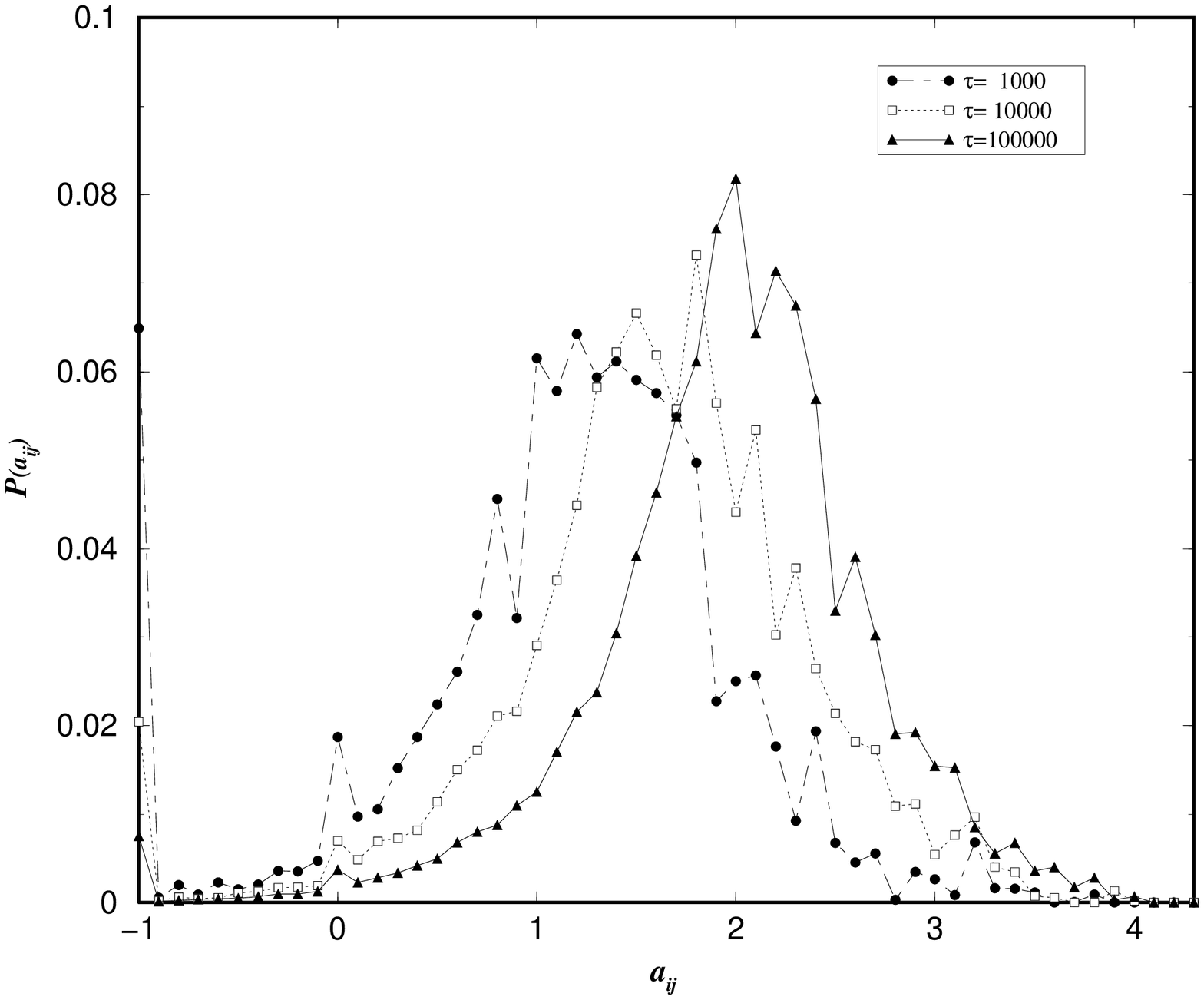}
  \caption{The vertical axis represents the frequency,
	   and the horizontal axis represents the values of the
	   off-diagonal elements of the matrix $(a_{ij})$ produced
	   under the {\it local\/} rule. The distribution of the elements
		   of this interaction matrix shifts toward the positive direction as time
		   increases.  The distribution is averaged over 10 samples of
		   simulations. Here
	   $\delta=\exp(-11)$, $\zeta=\exp(-10)$, $\xi=\exp(-9)$,
	   $v=1.0$ and $T=900$.}
  \label{hindo} 
  \end{minipage}
  \hspace{0.04\textwidth}
  \begin{minipage}{0.48\textwidth}
  \includegraphics[width=\textwidth]{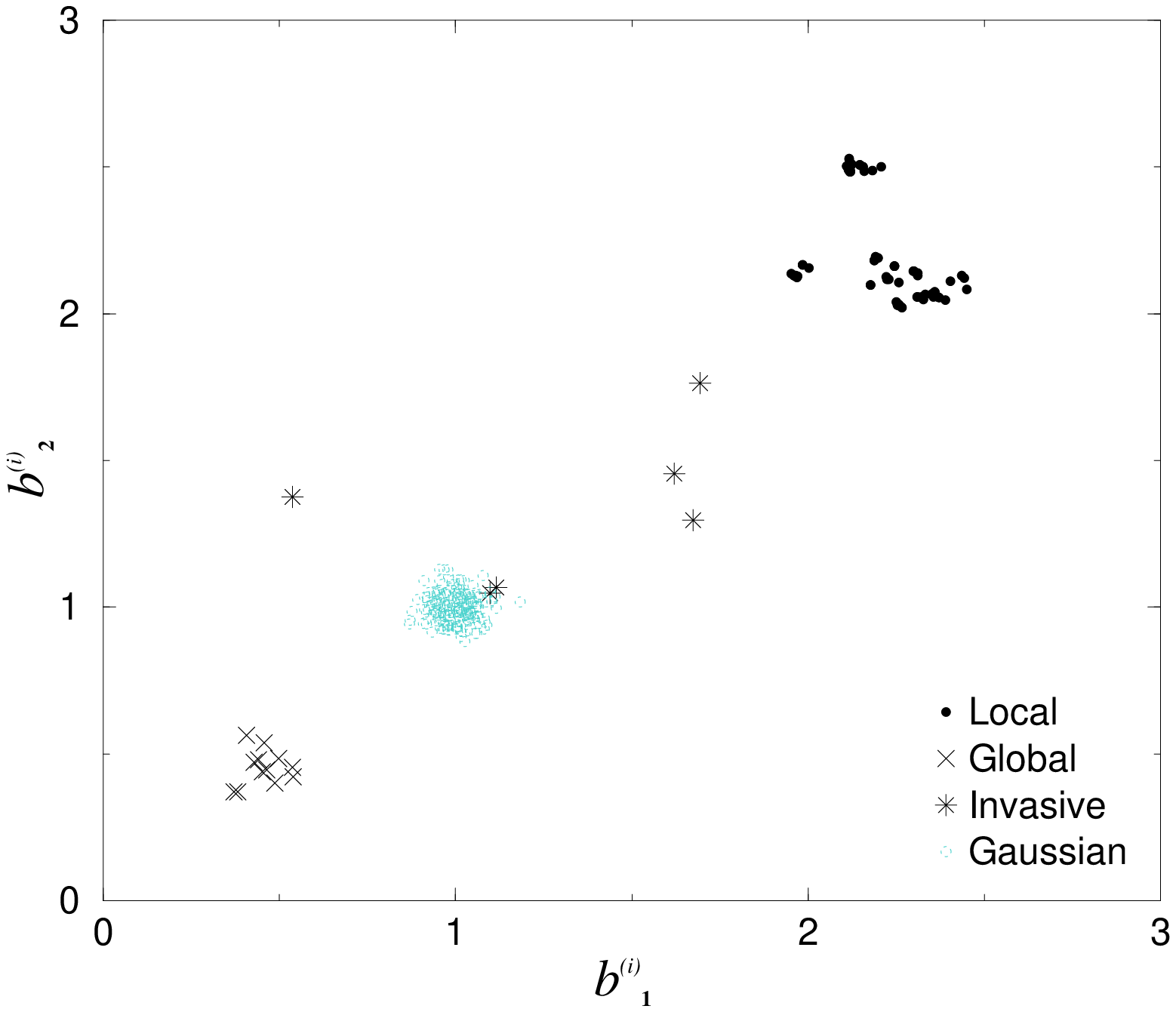}
  \caption{\label{bunpu} The distributions of the $\vecb{i}$ for
		   each species under {\it invasive}, {\it global\/}
	   and {\it local\/} rules at $\tau=10,000$. For reference,
		  the $\vecr{i}$ (described in the main text) also appear.
       As seen here, the $\vecb{i}$ under the {\it
	   local\/} rule deviate greatly from the $\vecr{i}$. Here
		   $\delta=\exp(-11)$,
	   $\zeta=\exp(-10)$, $\xi=\exp(-9)$, $v=1.0$ and $T=900$.}
 \end{minipage}

 \vspace{5mm}
\end{figure}

Since a subsystem of the type described above has a higher degree of
mutualism than the {\it observable} system, we expect that the
ecosystem will become more mutualistic as the diversity increases.  Since
the average fitness $\bar{f}=\sum_{j,k=1}^{N(t)} a_{jk} x_j(t) x_k(t)$
is a kind of average of the interaction coefficients, we can regard this
average as an index of mutualism. As seen in Figure \ref{diversity}, it
continues to increase only under the {\it local} rule. Figure
\ref{hindo} exhibits the distribution of $\{ a_{ij}\}$ for an ecosystem
that has emerged under the {\it local\/} rule. It should be noted that
these values are `naturally selected' from Gaussian random numbers with
mean $0$ and variance $1$. As seen in the figure, the average of the
distribution shifts to the positive area. This suggests that the
relationships between species tend to become more and more mutualistic
through the struggle for survival. 

Let us consider in detail the process of speciation by introducing a
$2S(\tau)$-dimensional {\it trait space}, where each species $i$ is
characterized by the point $(a_{i1},a_{i2},\ldots
,a_{i,S(\tau)},a_{1i},a_{2i},\ldots ,a_{S(\tau),i})$. In order to
visualize the distribution of the species in this high-dimensional
space, we project this point onto a point in a two-dimensional space:
\begin{equation}
 \vecb{i}
 = (\vbe{i}{1},\vbe{i}{2})\equiv
    \frac{1}{S(\tau)}
    \left(
      \sqrt{\sum_k a_{ki}^2}_,\quad
      \sqrt{\sum_j a_{ij}^2}
    \right)_. 
\end{equation} 
If the ecological status of
two species $k$ and $l$ are similar, $\vecb{k}$ and
$\vecb{l}$ are close to each other.

Figure \ref{bunpu} displays the sets $\{\vecb{i}\}$ for each species
with the interaction matrices resulting from {\it invasive\/}, {\it
global\/} and {\it local\/} rules. For the sake of contrast, we
also display the points $\vecr{i}$, defined in the same way as the
$\vecb{i}$, for a $50 \times 50$ matrix whose off-diagonal elements are
given as Gaussian random numbers with mean $0$ and variance $1$. We can
see that the sets $\{\vecb{i}\}$ for these three interaction matrices
differ greatly from $\{\vecr{i}\}$. This suggests that it is quite
inappropriate to use a random matrix to approximate an interaction
matrix of a real ecosystem.  Among these three sets $\{\vecb{i}\}$, that
obtained from the {\it local\/} rule in particular is shifted in
relation to $\{\vecr{i}\}$ more mutualistic region in the {\it trait
space}. This is suggested also by Figure \ref{hindo}. Moreover, several
groups of species are
observed in this case, indicating that the speciation processes under
the {\it local} rule do not result in a simple random drift or a random
diffusion in the trait space; that is, the groups of species
that appear in this case are fixed by the pressure of natural
selection.

\begin{figure}[tp]
 \begin{center}
  \includegraphics[width=0.48\textwidth]{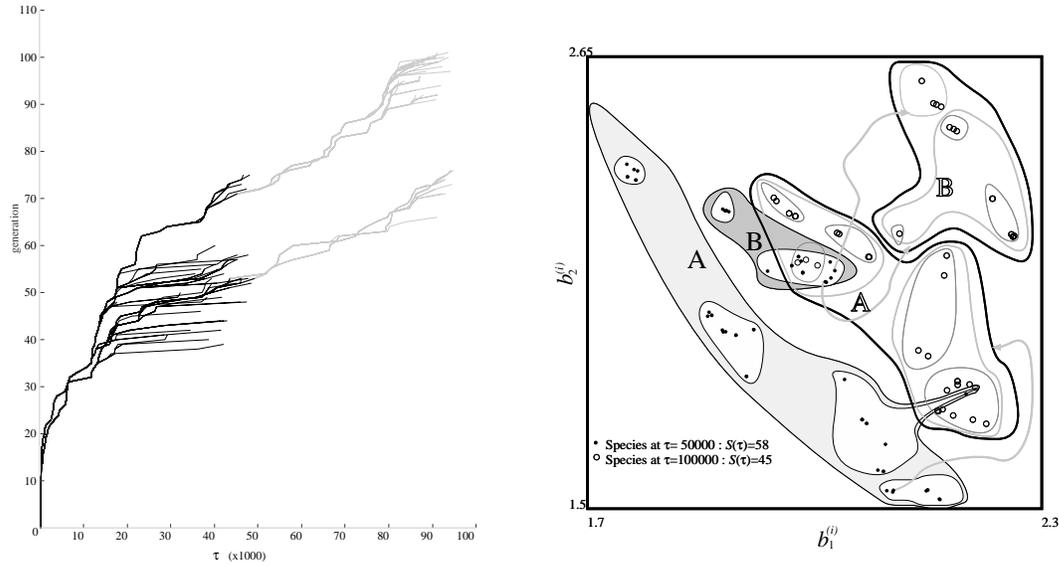}
  \hspace{0.02\textwidth}
  \includegraphics[width=0.48\textwidth]{TokitaYasutomiFig9b.eps}
\caption{\label{bunpu_local}(a): The genealogy of surviving
	   species at $\tau=50,000$ and $100,000$. Each species is plotted
	   as  a point whose location is determined by the time that this
		   species emerged (horizontal axis) and the generation that
		   this species represents (vertical line). 
	   Mother species and their daughter species are connected
	   by lines. The connections between existing species at
	   $\tau=50,000$ and their ancestors appear as
	   black lines, and those at $\tau =100,000$ as gray lines.
	   (b): This figure shows the following; (1) species closely related
	   in lineage possess similar characteristics (at least to the
		   extent that these are captured by the quantity plotted here); (2)
	   there is a hierarchical structure reminiscent of a 
	   family-genus-species; and (3) almost all species
	   existing at $\tau=50,000$ are extinct at $\tau=100,000$, and the
	   descendants of only three species survive and produce a large
		   number of descendents (adaptive radiation). Black circles indicate
	   $\{\vecb{i}\}$ at $\tau=50,000$ and white circles at
	   $\tau=100,000$. Here
	   $\delta=\exp(-11)$, $\zeta=\exp(-10)$, $\xi=\exp(-9)$,
	   $v=1.0$ and $T=900$. Species which belong to a given branch are
		   enclosed within the same closed curve. The thin closed
		   curves correspond to new branches, and the thick closed
		   curves correspond to old branches. Thus the thin curves
		   enclose a single genus and the thick curves enclose a
		   family.}
 \end{center}
\vspace{5mm}
\end{figure}

Figure \ref{bunpu_local} displays the relationship between species in
the case of the {\it local} rule. It is observed that this ecosystem
starts from a single species, and the diversity increases through
speciation. Figure \ref{bunpu_local}(a) displays the history of the
speciation. The black lines indicate a genealogy of species existing at
$\tau=50,000$, and the gray lines at $\tau=100,000$ from their ancestors
at $\tau=0$ and $\tau=50,000$, respectively. Lines representing the extinct species
before $\tau=50,000$ and $\tau=100,000$ are not plotted, although there are in
fact more branches for the extinct species. At $\tau=50,000$ there exist
58 species, and they are separated into two large groups which branches
at a very early stage.

Figure \ref{bunpu_local}(b) displays the points $\vecb{i}$ for each
species. The set $\{\vecb{i}\}$ at $\tau=50,000$ is plotted with black dots
and those at $\tau=100,000$ as circles.  We observe
two large groups in this figure: Groups A and B contain 39 and 19
species at $\tau=50,000$, respectively. Those species on the same
branch of the genealogy are plotted near each other in the
2-dimensional space. We also observe several subgroups in each
group: Group A contains four subgroups and group B contains two. These
groups also reflect the structure of genealogy, as those species in the
same subgroup appear on the same sub-branch of the genealogy. If we
think of each dot and circle as a `species', then each subgroup can be thought
of as a `genus' and an entire group is a `family'.  The two families
A and B still exist at $\tau=100,000$, shifting toward the upper right. All species at
$\tau=50,000$ are extinct, and those descendants of only three species
survive at $\tau=100,000$. Each arrow denotes a radiation from one of these
three to a group or a subgroup at $\tau=100,000$.
Descendants of one species of family A radiated into
$32$ species. This family separated into two genera, and even these
genera formed subsubgroups. There are $13$ species of descendants that
emerged from two
species of family B: one of these genera contains $4$ species and the other 9
species. The latter genus is separated into $4$ sub-genera.

\begin{figure}[tp]
 \begin{minipage}{0.6\textwidth}
  \includegraphics[angle=-90,width=\textwidth]{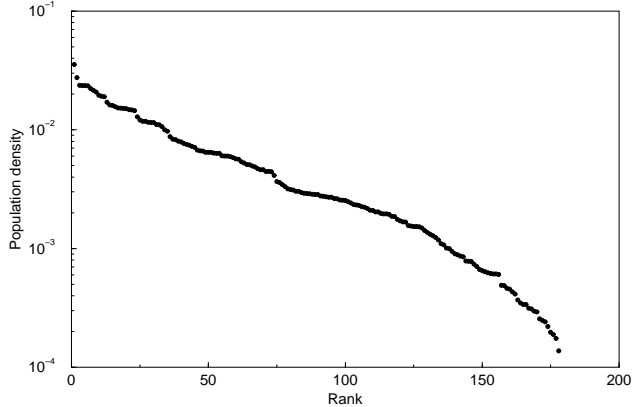}
 \end{minipage}
 \hspace{0.02\textwidth}
 \begin{minipage}{0.38\textwidth}
  \vspace{15mm}
  \caption{\label{xi2rank}Rank-size plot of the population
densities $x_i$ for $178$ observable species at $\tau=150,000$. }
 \end{minipage}
\end{figure}

Finally, we note the interesting correspondence of a result
obtained in the present study to a well-known ecological law of
population abundance: As seen from Figure \ref{xi2rank}, the
dominance-diversity distribution is $S$-shaped, and its middle part
displays a power law behavior. Such behavior is observed in many species-rich
communities (Hubbell, 1997; Magurran, 1988).
This suggests that the present model shares some
characteristics with populations in nature. Let us here note that our
data were not obtained through any averaging, but from only one
simulation. This is significant because in order to obtain such a
distribution using one sample, it is required that the sample produce
a sufficiently large number of species. This has been realized
for the first time in the present study.

%%
%% Referee #2 [1]2ndに従い追加(2002/10/17)
%%
In the present study, we have mentioned ``invasion type'' mutants with
respect to species-specific interaction by the {\it local}
rule. However we can also think of "mutant type" mutants with respect
to species-specific interaction, which have the parameter deviated
from the parent's value by a random quantity. We have executed similar
simulations by such a {\it mutative local} rule and found a similar
diversification like the {\it local} rule. Under the {\it mutative
local} rule, the diversity $S(\tau)$ increases little slower but is fitted
by a power function like under the {\it local} rule.

{}From the above discussion, we conclude that an ecological
system which is composed of many hierarchically structured species
strongly interacting with each other {\it can\/} emerge through an
evolutionary speciation process by the {\it local} rule.

\section{Resistance to Invasion}

\subsection{Method}

As mentioned in the first section, Elton (1958) suggested that complex
communities are more stable than simple ones. In this context, as
pointed out by Case (1990, 1991), Elton's definition of the `stability' is
not the asymptotic stability of the assembled matrix but the 
resistance to invasion of communities. We thus can understand his
hypothesis as an assertion that species-rich ecosystems are more
resistant to invasion by exotics than are species-poor ones.

Case (1990, 1991) constructed models to check Elton's
hypothesis from this point of view. He randomly assembled
ecosystems and selected stable and feasible ones.  Then he tested
their resistance to invasion by adding new species generated by a rule
very similar to our {\it invasive} rule. He concluded that a more
diverse and more strongly connected ecosystem has greater 
resistance to invasion.

\subsection{Results}

We carried out a similar test on ecosystems formed under the
{\it global} and {\it local} rules and compared the 
resistance to invasion of these ecosystems. There is an important
difference between these two rules and the {\it
invasive\/} rule. Under the {\it global} and {\it local}
rules, new species are mutants of the existing species.
However, under the {\it invasive\/} rule these are independent of
the existing species, invading from without. In our tests, we added mutants
generated by the {\it local\/} and {\it global\/} rules for only the
first $R$ periods, and then we let invaders
generated by the {\it invasive\/} rule attack the ecosystem for 200
periods. $R$ denotes a kind of maturity index of the ecosystem.

\begin{figure}[tp]
 \begin{minipage}{0.6\textwidth}
  \includegraphics[angle=-90,width=\textwidth]{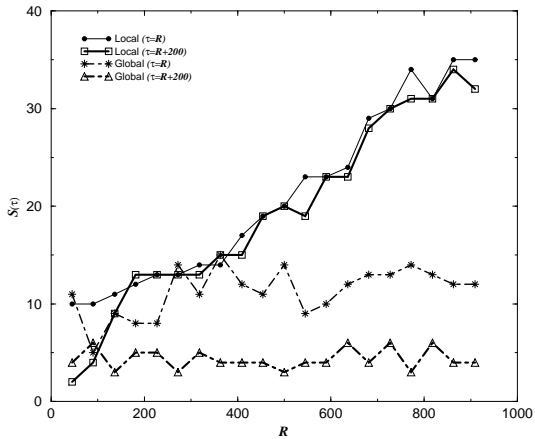}
 \end{minipage}
 \hspace{0.01\textwidth}
 \begin{minipage}{0.39\textwidth}
  \caption{\label{meneki}After $R$ periods of assemblage
	   under {\it local} and {\it global} rules, invaders
	   generated by the {\it invasive} rule attack for 200
	   periods. When $R$ is small, the invaders
	   succeed in destroying the existing ecosystem. However, When $R$ is
	   larger than 100, the ecosystem developed under {\it
	   local} rule can maintain its diversity. Contrastingly, that
		   developed under the {\it global} rule experiences a
		   reduction in diversity resulting from attacks by invaders.
	   Here $\delta=\exp(-11)$, $\zeta=\exp(-10)$, $\xi=\exp(-9)$,
	   $v=1.0$ and $T=900$.}
 \end{minipage}
\vspace{5mm}
\end{figure}

The results of our experiment are depicted in Figure \ref{meneki}, which
shows the dependence of the resulting diversity $S(\tau)$ on $R$.
As seen there, when $R$ is small, the diversity is drastically 
reduced under either rule.  However, when $R$ is larger than $200$, ecosystems
developed under the {\it local\/} rule maintain their diversity, resisting
invasion. Contrastingly, those developed under the {\it global\/}
rule exhibit a reduction in
their diversity for any value of $R$, as a result of destructive
invasions.  This result suggests that an ecosystem developed under the {\it
local\/} rule has strong resistance to invasion if it is sufficiently
mature.

The difference between the {\it local\/} and {\it global} rules results
from the difference in their levels of mutualisms. As we have seen in
the previous section, an ecosystem evolving under the {\it local\/}
rule is more mutualistic than one evolving under the {\it global\/} rule. We
should note that the average fitness $\bar{f}$ here again turns out to
play an important role as an index indicating the resistance to invasion . If an
ecosystem has an average fitness $\bar{f}$, the fitness $f_I$ of an invader
 must be larger than $\bar{f}$ in order for the invader to succeed.
Since $f_I$ can be approximated by a Gaussian random number with
mean $0$ and variance $1$, for an ecosystem with sufficiently large
diversity, the probability for a successful invasion is given by the {\it
error function},
\begin{equation}
{\rm Prob.}(f_I>\bar{f}) = {\rm Erfc}(\bar{f}) 
 = \frac{1}{\sqrt{2\pi}}\int_{\bar{f}}^\infty 
   {\rm exp}\left(-\frac{t^2}{2}\right){\rm d}t,
\label{Errfnc}
\end{equation}
which decreases exponentially as $\bar{f}$ increases.  From
Figure \ref{diversity}(b), it is possible to estimate the probability 
(\ref{Errfnc}) by using the value of $\bar{f}$ for
$R\sim 200$ at which the difference between the {\it local} and {\it
global} becomes larger in Figure \ref{meneki}.  The average
fitness for the {\it local} rule, $\bar{f}(\tau=200)\sim 1$, gives ${\rm
Erfc}(1)\sim 0.16$, while that for the {\it global} rule,
$\bar{f}(\tau=200)\sim 0.25$, gives ${\rm Erfc}(0.25)\sim 0.4$.  Figure
\ref{diversity}(b) indicates that $\bar{f}$ is a monotonically increasing
function of $R$ for the {\it local} rule, and hence we expect that a
random invasion becomes almost impossible against a fully-matured
ecosystem (i.e., one of large $R$) under the {\it local} rule. That is, only
temporally neutral mutants can influence or create extinction events in an existing
ecosystem evolved under the {\it local} rule.

\section{Discussion}

\subsection*{Generalized Lotka-Volterra equations vs. replicator equations}

The generalized Lotka-Volterra equations (GLVE) are more commonly used than
the replicator equations (RE) in the field of mathematical biology.
However, it is known that an $N$-dimensional RE is mathematically
identical to the ($N-1$)-dimensional GLVE (Hofbauer \& Sigmund, 1998)
\begin{equation}
 \frac{{\rm d}y_i}{{\rm d}t} = y_i (r_i + \sum_{j=1}^{N-1}b_{ij}y_j) \qquad 
(i=1, 2, \cdots , N-1), \label{LVeq}
\end{equation}
where the $i$th species' population $y_i$ is defined in terms of the
population in the RE (\ref{replicator_eq}), $x_i$ by 
\begin{equation}
y_i = \frac{x_i}{x_N}. \label{transyi}
\end{equation}
The value of $y_i$ can take any non-negative real
number. The interactions $\{b_{ij}\}$ and
the intrinsic growth (or decay) rate $r_i$ here are defined in terms of
interactions in the RE by 
\begin{eqnarray}
b_{ij}&=&\left.a_{ij}-a_{Nj}\right._,\label{transbij}\\
r_i&=&\left.a_{iN}-a_{NN}=a_{iN}+v\right._.\label{transri}
\end{eqnarray}
  Any orbit of the GLVE can therefore be mapped onto
an equivalent orbit of the RE, and any behavior exhibited by these two
systems can be studied using either, at least in principle.

Here we explain why we nevertheless study the RE rather than the GLVE.
As described in the first section, Taylor (1988b) pointed out that most
of the dynamics resulted in explosions of populations as proportion of
species with $r_i>0$ increased or proportion of positive $b_{ij}$
increased.  Moreover, the populations of species with low fitness often
take extremely small values that cause underflow in naive numerical
computational scheme. Consequently, numerical analysis of the GLVE often
encounters both divergence and underflow, in particular for a system
with high diversity and complex interactions. On the other hand, in the
RE, population explosions are avoided by definition, because any orbit
of the population $\{x_i\}$ is bound in the simplex
$\sum_{i=1}^N x_i=1$. Moreover, in the present model, the problem of
underflow is also avoided because the heteroclinic orbits, the cause of
underflow, are excluded by the introduction of the {\it extinction
threshold} (Tokita \& Yasutomi 1999). These are the reasons we
use the RE rather than the GLVE.

\begin{figure}[tp]
 \begin{minipage}{0.55\textwidth}
  \includegraphics[angle=-90,width=\textwidth]{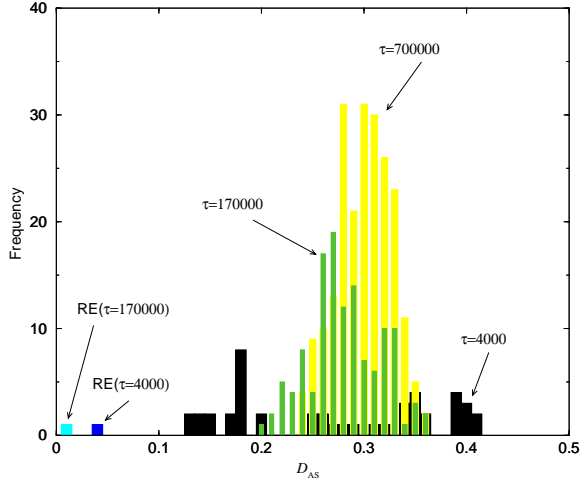}
 \end{minipage}
 \hspace{0.02\textwidth}
 \begin{minipage}{0.43\textwidth}
  \caption{\label{DasHist}The vertical axis represents the
	   frequency, and the horizontal axis represents the values of the
	   antisymmetricity $D_{AS}(B)$ at $\tau=4,000, 170,000$ and $700,000$ and
           $D_{AS}(A)$ at $\tau=4,000$ and
	   $170,000$ (``RE''). The value of the antisymmetry 
	   $D_{AS}(A)$ at $\tau=700,000$ becomes nearly zero. Here
	   $\delta=\exp(-11)$, $\zeta=\exp(-10)$, $\xi=\exp(-9)$, $v=1.0$ and
	   $T=1,000$.}
 \end{minipage}
\vspace{5mm}
\end{figure}

Let us note that the meaning of the interaction coefficients
$a_{ij}$ in the RE is different from that of the $b_{ij}$ in the GLVE:
even if the $a_{ij}$ and $a_{ji}$ take only positive values (mutualistic
relationships), the $b_{ij}$ and $b_{ji}$ can take negative
values (competitive or exploitative relationships) depending on the
values of $a_{ij}$ and $a_{Nj}$ in Eq. (\ref{transbij}). 
%
% 以下新しく加えたところ
As the growth rate of the species $i$ is defined by the fitness
$\sum_ja_{ij}x_j$ deducted by its average $\sum_{j,k}a_{jk}x_jx_k$ in
the definition of the replicator equations (\ref{replicator_eq}), the
ecological significance of the developed ecosystem in the present study
should be discussed by the matrix $B=(b_{ij})$ in the GLVE than
$A=(a_{ij})$ in the RE. We therefore study the nature of matrices $B$
transformed from $A$ by Eq. (\ref{transbij}). Here we should note that
there are $N$ ways of transformation from $A$ to $B$ because of the
arbitrary option of the species $N$ in Eq. (\ref{transyi}). To
characterize a $L\times L$ matrix $M=\{m_{ij}\}$, here we define ``antisymmetricity'' as
\begin{equation}
D_{AS}(M)\equiv \left.\frac{\langle (m_{ij}-m_{ji})^2\rangle_M}{4\langle m_{ij}^2\rangle_M}\right._.
\end{equation}
The bracket $\langle (\cdots )\rangle_M\equiv
\frac{1}{L(L-1)}\sum_{i=1}^L\sum_{j\neq i}^L(\cdots )$ denotes the
average over all off-diagonal elements of $M$. For the RE with
interaction $A$ at $\tau$, $L$ becomes $S(\tau)$, and $L=S(\tau)-1$ for
$B$ of the corresponding GLVE.  In general, the antisymmetricity
$D_{AS}(M)$ becomes $0$, $1/2$ and $1$ when $M$ is symmetric
($m_{ij}=m_{ji}$), randomly asymmetric ($m_{ij}\neq m_{ji}$) and
antisymmetric ($m_{ij}=-m_{ji}$), respectively. Figure \ref{DasHist}
plots the distribution of $D_{AS}(B)$ at $\tau=4,000, 170,000 \mbox{ and
} 700,000$, where $S(4,000)=45, S(170,000)=125 \mbox{ and }
S(700,000)=216$. The value of the antisymmetry $D_{AS}(A)$ of the RE at
$\tau=4,000, 170,000$ is also indicated.  From the figure, the
interaction matrix $A$ of the RE turns out to approach the symmetric
point ($D_{AS}=0$; mutualism). On the other hand, the matrix
$B$ tends to approach the intermediate region ($D_{AS}\sim 0.3$) between
symmetric and asymmetric. This implies that the system evolved under the
{\it local} rule has complex ecological interspecies interactions
including mutualism, competition, predation and parasitism.

From the figure \ref{hindo} and \ref{DasHist}, we conclude that the
frequency of the interspecies interactions of the emerged system
can be approximated by a symmetric Gauss distribution with mean $m(>0)$
and variance $1$
\begin{eqnarray}
 P(a_{ij})&=&\frac{1}{\sqrt{2}}\exp\left(-\frac{(a_{ij}-m)^2}{2}\right),
   \quad (\mbox{for}\quad i\neq j \quad\mbox{and}\quad a_{ij}=a_{ji})
   \label{distaij1}\\
 a_{ii}&=&-v(<0)\label{distaij2}
\end{eqnarray}
in the infinite limit of the number of species ($N\to\infty$), where
$m>0$ is the level of mutualism. This type
of the replicator equation, in general, has a number of saturated fixed
points, which is proved by the symmetry of the interaction (Hofbauer \&
Sigmund, 1998). 

Finally, let us see how \{$r_i$\} and \{$b_{ij}$\} in the
corresponding GLVE mutate in the {\it local} rule.  First, values
$r_k$, \{$b_{ik}$\} and \{$b_{ki}$\} for all $i$ of a new species $k$
are copied from $r_j$, \{$b_{ij}$\} and \{$b_{ji}$\} of a parent
species $j$, respectively. Second, by the definition of the {\it
local} rule and the transformation (\ref{transbij}) and
(\ref{transri}), $r_k$ or $b_{lk}$ (and $b_{kl}$) for arbitrary $l$ is
replaced by a random quantity. Note that $r_k$ is replaced by a random
number with mean $v$ and variance $1$ while $b_{lk}$ ($b_{kl}$) with
mean $0$ and variance $1$.  From the transformation (\ref{transbij})
and the distribution (\ref{distaij1}) and (\ref{distaij2}) of
$a_{ij}$, the distribution of $b_{ij}$ becomes a Gauss distribution
with mean $0$ and, therefore, is not mutualistic. Minor mutualist
species in the RE, therefore, are not necessarily mutualists in the
GLVE, and do not necessarily lead population explosion.  On the other
hand, $r_i$ in Eq. (\ref{transri}) is biased by $v(>0)$ and the mean
value of $r_i$ increases over time because mean value of $a_{iN}$
increases as indicated by Fig. \ref{hindo}.  Accordingly, in terms of
the GLVE, proportion of producers increases as time goes on in the
{\it local} rule. As pointed out by Taylor (1988b), this may trigger
population explosion.  Another chance for explosion is implied in the
transformation (\ref{transyi}) when the $N$-th ``base'' species
goes extinct ($x_N(t)\to 0$), which breaks down the simulation of the
corresponding GLVE.  We stress here that even if the extinction of one
``base'' species causes the explosion in the GLVE, the evolution in
the RE proceeds successfully and there are still other equivalent GLVE
systems by transformations by other $N-1$ ``base'' species
as $y_i=x_i/x_M$ ($M=1, 2, \cdots , N-1$).  The RE, therefore, traces
a possible path of evolution avoiding the breakdown of the simulation
which may occur in the GLVE. This methodological advantage
of the RE is related to the circumstance that the dimension (the
degree of freedom) of the RE is one higher than the GLVE
and the total population is conserved (Eq. (\ref{constraint})). This
suggests that such a conservative quantity, e.g. resource limitations,
would be essential in the modeling of multispecies evolution using the
GLVE.
\vspace{2cm} The authors would like to thank Yoh Iwasa, Takashi Ikegami and Tsuyoshi
Chawanya for their helpful comments on the manuscript
and for their encouragement. This work was partially supported by the
Japan Society for the Promotion of Science, a Grant-in-Aid from the
Ministry of Education, Science, Sports and Culture of Japan
(No. 12740235, 13831007 and 14740232), and the Suntory
Foundation. Some of the numerical calculations were carried out on a
Fujitsu VPP500/40 at the Supercomputer Center, Institute for Solid
State Physics, University of Tokyo and on the Machikaneyama PC Cluster
System in the Condensed Matter Theory Group, Graduate School of
Science, Osaka University ({\tt URL:
http://wwwacty.phys.sci.osaka-u.ac.jp/\~{}mhill}).

\end{document}